\documentclass[aps,pra,twocolumn,superscriptaddress,amsmath,amssymb,amsfonts,floatfix]{revtex4}

\usepackage{graphicx}

\begin{document}
\title{Quantum phase transitions of two-species bosons in square lattice}

\author{Pochung Chen}
\email{pcchen@phys.nthu.edu.tw} %
\affiliation{Department of Physics, National Tsing Hua University,
Hsinchu 30013, Taiwan}

\author{Min-Fong Yang}
%\email{mfyang@thu.edu.tw} %
\affiliation{Department of Physics, Tunghai University, Taichung
40704, Taiwan}

\date{\today}

%%%%%%%%%%%%%%%%%%%%%%%%%%%%%%%%%%%%%%%%%%%%%%%%%%%%%%%%%%%%%%%%%%%
\begin{abstract}
We investigate various quantum phase transitions of attractive
two-species bosons in a square lattice. Using the algorithm based
on the tensor product states, the phase boundaries of the pair
superfluid states with nonzero pair condensate density \emph{and}
vanishing atomic condensate density are determined. Various
quantum phase transitions across the phase boundaries are
characterized. Our work thus provides guides to the experimental
search of the pair superfluid phase in lattice boson systems.
\end{abstract}
%%%%%%%%%%%%%%%%%%%%%%%%%%%%%%%%%%%%%%%%%%%%%%%%%%%%%%%%%%%%%%%%%%%

         % Multicomponent condensates; spinor condensates

\maketitle

%%%%%%%%%%%%%%%%%%%%%%%%%%%%%%%%%%%%%%%%%%%%%%%%%%%%%%%%%%%%%%%%%%
%\section{introduction}

Ultracold atoms in optical lattices provide an unprecedented
toolbox for the experimental realization of exotic quantum phases
which are not easily accessible in condensed matter
systems~\cite{BEC_reviews}. Recent experimental demonstration of
the superfluid-to-Mott transition for ultracold bosons in an
optical lattice~\cite{BEC_exps}, for example, has paved the way
for studying other strongly correlated phases in various lattice
models. Among these systems, Bose-Bose, Fermi-Fermi, and
Bose-Fermi mixtures have attracted considerable attention, since a
variety of interesting ordered states can be realized in such
systems~\cite{BEC_reviews}.
In particular, it has been suggested that, for the mixture of two
bosonic species in a lattice, interspecies repulsion and
attraction can give rise to additional incompressible
super-counterfluid (SCF) and compressible pair superfluid (PSF)
phases, respectively~\cite{kuklov03,altman03,Isacsson05,Powell09,soyler09,hubener09,%
kuklov04-1,kuklov04-2,Guertler08,arguelles07,mathey09,hu09,Menotti10,Iskin2010,Trefzger09}.
Furthermore, when the intra-species nearest-neighbor repulsion is
included, the counterflow supersolid and the pair-supersolid
phases may also emerge~\cite{Trefzger09}.

The quantum phase of our interest is the PSF phase of two-species
hard-core bosons on a square lattice with interspecies attraction
$U_{ab}(<0)$, which is described by the Hamiltonian,
\begin{equation}
H= - t \sum_{\langle ij \rangle} \left( a_i^\dag a_j + b_i^\dag b_j + \hbox{h.c.} \right)
   -  \mu \sum_{i,\sigma} n_i^\sigma  + U_{ab} \sum_i n_i^a n_i^b  , \label{eqn:H}
\end{equation}
where $\sigma=a,b$ indicates the two species and $n_i^\sigma$ is
the particle densities at site $i$. The intra-species tunneling is
denoted by $t>0$ and $\mu$ is the chemical potential. The symbol
$\langle ij \rangle$ indicates nearest neighbors. When $|U_{ab}|/t
\gg 1$, the PSF phase can be stabilized by forming a superfluid of
boson pairs of different
species~\cite{kuklov04-1,kuklov04-2,Guertler08,arguelles07,mathey09,hu09,Menotti10,Iskin2010,Trefzger09}.
This phase is characterized by a nonzero pair condensate density
$n_0^{\text{PSF}} \equiv |\langle ab \rangle|^2 \neq 0$ \emph{and}
a vanishing atomic condensate densities $n_0^\sigma \equiv
|\langle \sigma\rangle|^2 = 0$. By considering two species as two
components of spin-1/2 particles, it implies from the symmetry
breaking perspective that the ``charge" $U(1)$ symmetry
$a\rightarrow e^{i\theta}\, a$; $b\rightarrow e^{i\theta}\, b$ is
broken spontaneously in the PSF phase, while the ``spin" $U(1)$
symmetry $a\rightarrow e^{i\varphi}\, a$; $b\rightarrow
e^{-i\varphi}\, b$ is not. Here $\theta$ and $\varphi$ denote the
angles of the corresponding $U(1)$ rotations. The PSF phase hence
can also be identified by a nonzero pair-superfluid density
$\rho_{\rm s}^{\text{PSF}} \propto \langle (W_a+W_b)^2\rangle \neq
0$ \emph{and} a zero super-counterfluid density $\rho_{\rm
s}^{\text{SCF}} \propto \langle (W_a-W_b)^2\rangle=0$, where
$W_{\sigma}$ are the winding numbers for bosons of species
$\sigma$~\cite{Pollock87}. By contrast, the two miscible
superfluid (2SF) phase is expected in the opposite limit of
$|U_{ab}|/t \ll 1$, where both species form superfluids with
nonzero atomic condensate density $n_0^a$, $n_0^b \neq 0$. In
addition to the distinct properties of the PSF states, intriguing
quantum phase transitions out of this phase can occur. Based on
the analysis of the mean-field theory, it was proposed
that~\cite{kuklov04-2}, for the cases of spatial dimensions being
larger than one, the PSF-2SF transitions can be either second or
first order, while the PSF to Mott insulating (MI) phase
transitions are always second order.

While the existence of a PSF phase of lattice boson has been
pointed out in the literature using various
approaches~\cite{kuklov04-1,kuklov04-2,Guertler08,arguelles07,mathey09,hu09,Menotti10,Iskin2010,Trefzger09},
direct calculations of microscopic quantum models with large sizes
has not yet been reported except for the one-dimensional
case~\cite{arguelles07,mathey09,hu09}. This is perhaps due to the
difficulty in analyzing these quantum models by applying
conventional methods. For example, despite quantum Monte Carlo
(QMC) simulation being free from the sign problem for this model,
several practical technical issues may still hinder high-precision
large-scale QMC calculations. Firstly, multi-worm algorithm needs
to be implemented~\cite{Guertler08,soyler09,Pollet}, because
standard single-worm algorithm may converge quite slowly in the
PSF states. Secondly, very low temperatures are necessary to
detect the pair superfluid coherence in the PSF phase. This is due
to that the effective hopping of \emph{pairs} is about the order
of $t^2/|U_{ab}|$~\cite{kuklov03,altman03} and is quite small
since $t/|U_{ab}| \ll 1$ in the PSF phase. Finally, systems of
large enough sizes should be considered in order to determine
phase transitions being of either \emph{weakly} first order or
second order. Therefore, accurate quantitative predictions of
associated quantum phase transitions in the thermodynamical limit
around the PSF phase remain under investigation.

\begin{figure}[tb]
\includegraphics[width=0.9\columnwidth]{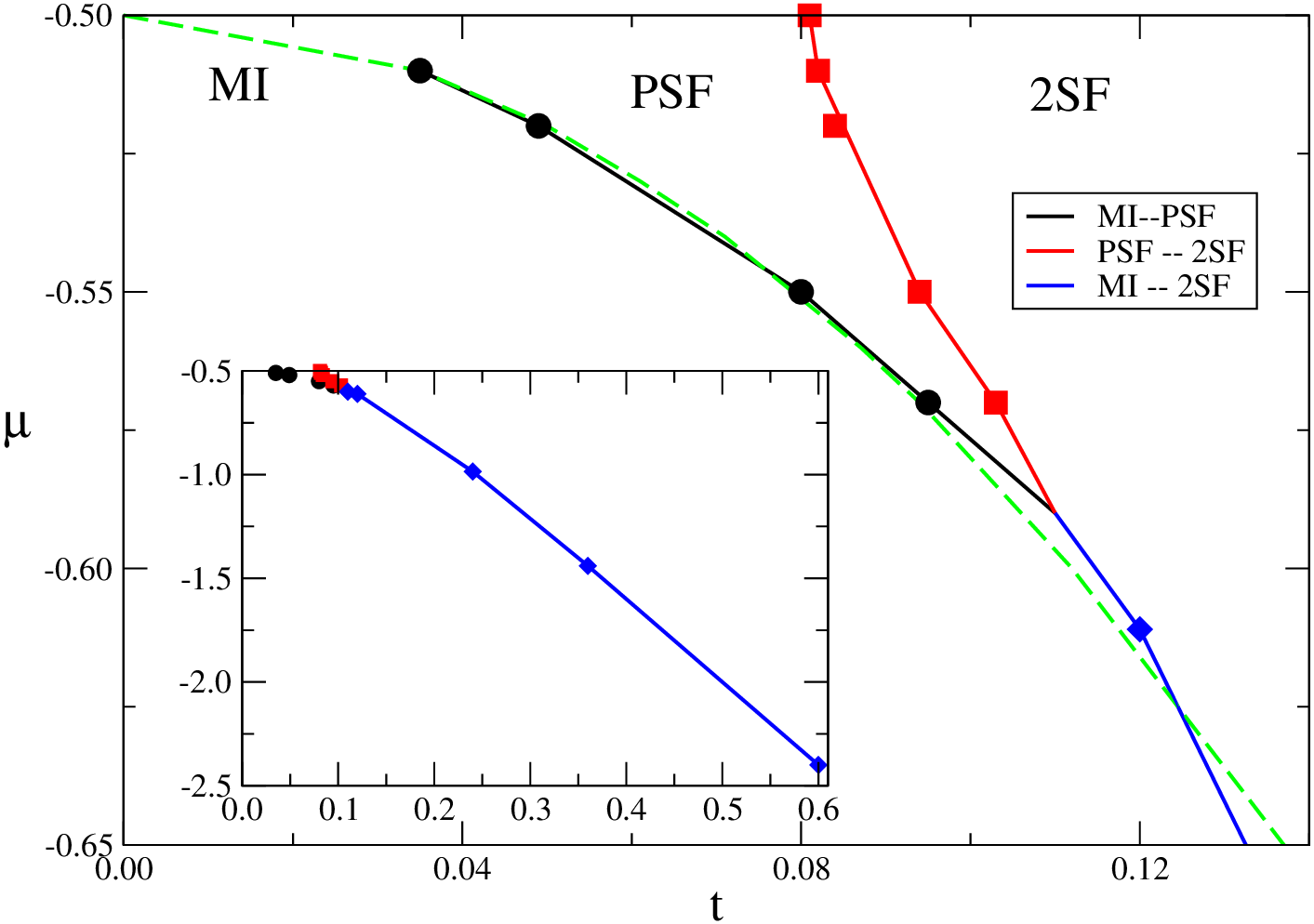}
\caption{(Color online)
Ground-state phase diagram of attractive two-species hard-core
bosons on a square lattice of size $2^7\times 2^7$ as described in
Eq.~\eqref{eqn:H} with $|U_{ab}|=1$. Solid lines are guide to eyes.
The green dashed line gives the analytic estimate of the MI-PSF
phase boundary (see text).
The inset shows the phase diagram in a larger scale. }
\label{fig:phase_diag}
\end{figure}

In this paper, various quantum phase transitions of the model in
Eq.~\eqref{eqn:H} are investigated. We would set $|U_{ab}| \equiv
1$ as the energy unit. Due to the particle-hole symmetry in the
present hard-core model, the phase diagram is symmetric with
respect to the line $\mu=-1/2$, at which the average density for
each species becomes $1/2$. It is therefore sufficient to
concentrate on the parameter region of $\mu \leq -1/2$ where both
species are below half filling.
Our main purpose is to determine quantitatively the phase
boundaries of the PSF phase and to reveal the nature of these
quantum phase transitions. Here the combined
algorithm~\cite{iTEBD-TRG} of the infinite time-evolving block
decimation (iTEBD) method~\cite{iTEBD} and the tensor
renormalization group (TRG) approach~\cite{TRG} is exploited. The
success of this method in obtaining accurate physical quantities
at zero temperature of systems with large sizes has been
demonstrated for several quantum spin
systems~\cite{iTEBD-TRG,Chen09,Chen10,Zhao10,Li10}. Since systems
of two-species hard-core bosons can be mapped onto anisotropic
spin-1/2 bilayer spin models, from our previous experience on the
study of the frustrated bilayer spin models~\cite{Chen10}, the
employed approach is expected to give reliable results for the
present issues. The resulting phase diagram is shown in
Fig.~\ref{fig:phase_diag}. The region of the PSF phase with finite
pair condensate but zero atomic condensate is discovered, and the
phase boundaries out of the PSF state are determined precisely.
The nature of various transitions is found to be in agreement with
that proposed in Ref.~\cite{kuklov04-2}. Our findings hence provide
further theoretical support on searching the PSF phases in the
bosonic systems with mutual attraction. However, we stress that
the PSF phase is stabilized only within a limited region of the
zero-temperature phase diagram in Fig.~\ref{fig:phase_diag}.
Therefore, carefully tuning system parameters into the suggested
parameter regime is necessary to uncover experimentally this novel
phase.

%%%%%%%%%%%%%%%%%%%%%%%%%%%%%%%%%%%%%%%%%%%%%%%%
%%%%%%%%%%%%%%%%%%%%%%%%%%%%%%%%%%%%%%%%%%%%%%%%

%%%%%%%%%%%%%%%%%%%%%%%%%%%%%%%%%%%%%%%%%%%%%%%%%%%%%%%%%%%%%%%%%%
%\section{model and method}

To characterize different phases, several local order parameters
are evaluated. Because the hopping parameter and the chemical
potential in Eq.~\eqref{eqn:H} are set to be equal for both
species, equal densities for the two species ($n^a = n^b$) are
expected. In our calculations we do find that the exchanging
symmetry between two species is always unbroken and the
expectation values for $a$ and $b$ bosons are the same.
Consequently, we show only the results of species $a$ as
representatives. When the average density of single species boson
$n = (1/N_\textrm{s}) \sum_i \langle n_i^a \rangle$ becomes an
integer (here $N_\textrm{s}$ denotes the number of lattice sites),
systems are in the MI states. Otherwise, systems belong to either
the PSF or the 2SF phases. As mentioned before, the last two
phases can be distinguished by the expectation values of the
atomic and the pair condensate density $n_0=|\langle a\rangle|^2$
and $n_0^{\text{PSF}}=|\langle ab\rangle|^2$, respectively. Here
$\langle a \rangle = (1/N_\textrm{s}) \sum_i \langle a_i \rangle$
and $\langle ab \rangle = (1/N_\textrm{s}) \sum_i \langle a_i b_i
\rangle$. In the 2SF phase, one has $n_0\neq 0$, while in the PSF
phase one has $n_0=0$ and $n_0^{\text{PSF}}\neq 0$.

%%%%%%%%%%%%%%%%%%%%%%%%%%%%%%%%%%%%%%%%%%%%%%%%%%%%%%%%%%%%%%%%%%
%\section{tensor network algorithm}

\begin{figure}[tb]
\includegraphics[clip,width=0.9\columnwidth,height=0.5\columnwidth]{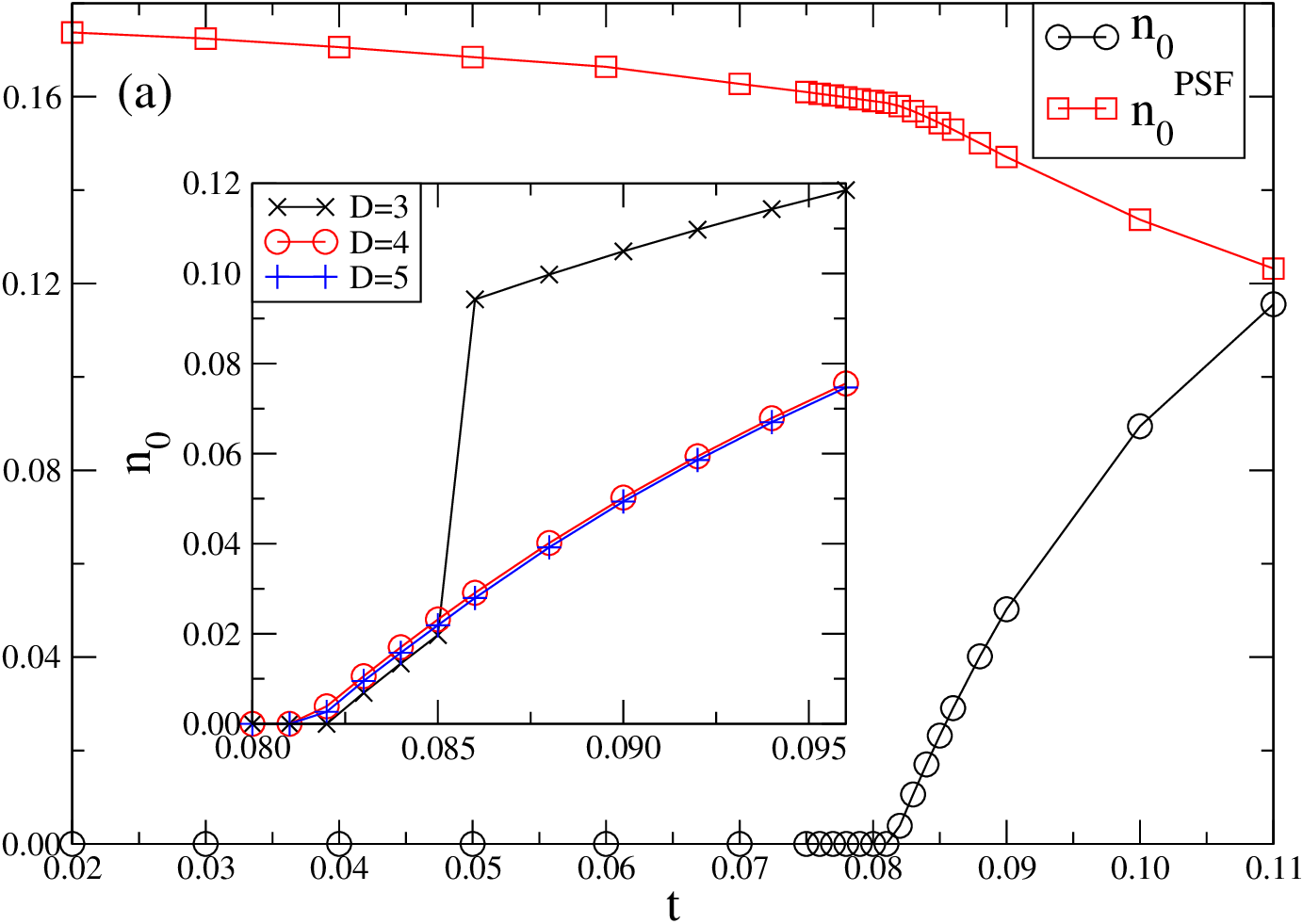}
\vskip 0.2cm
\includegraphics[clip,width=0.9\columnwidth,height=0.5\columnwidth]{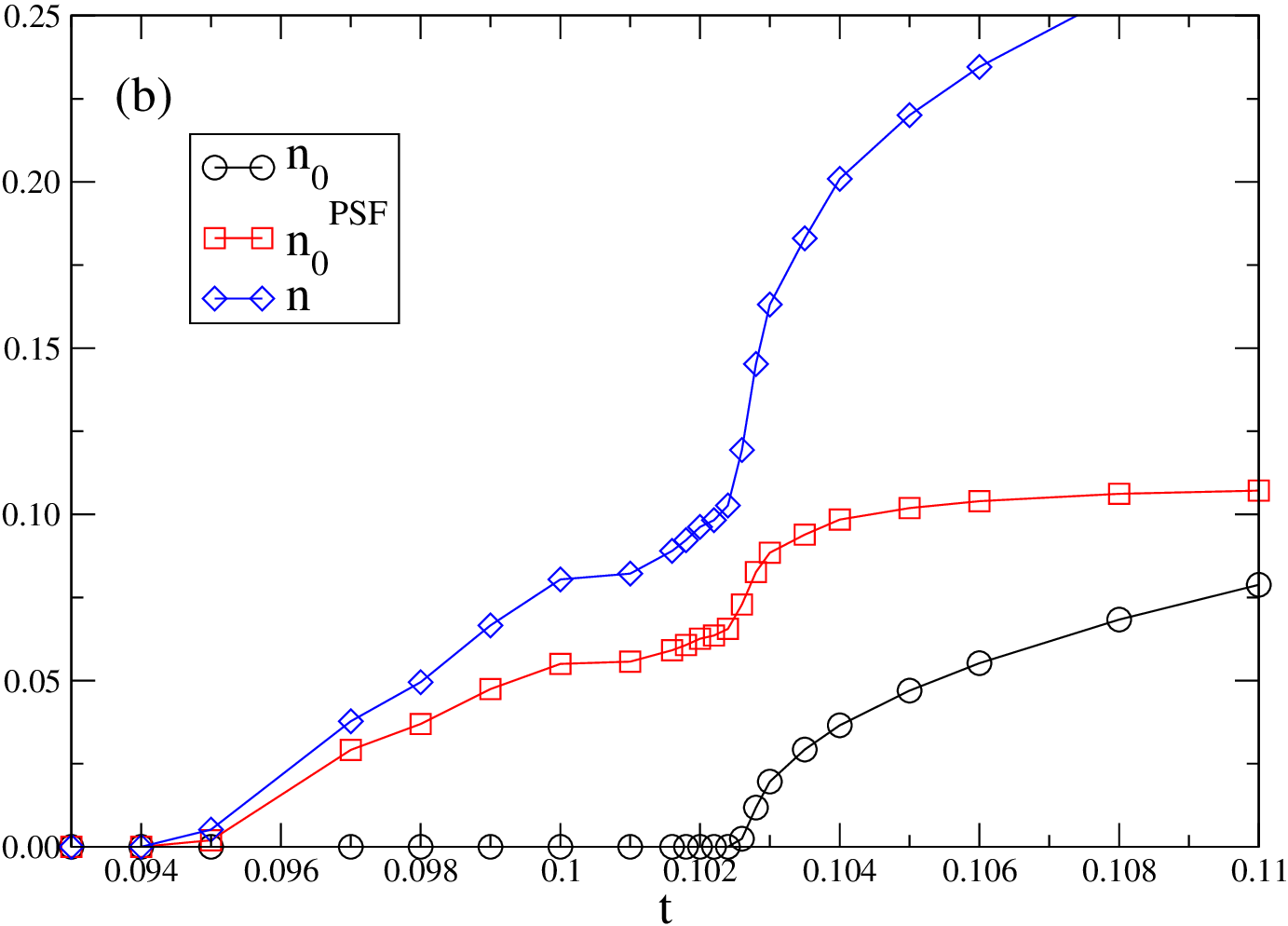}
\caption{(Color online) Values of particle density $n$, atomic
condensate density $n_0$, and pair condensate density
$n_0^\textrm{PSF}$ for the ground states at (a) $\mu=-0.5$ and (b)
$-0.57$ as functions of hopping parameter $t$. The inset in (a)
illustrates the $D$ dependence of $n_0$, where we use $D_{\rm
cut}=16$ for $D=3$ and $D_{\rm cut}=25$ for $D=4$, 5. }
\label{fig:scan3-4}
\end{figure}

\begin{figure}[tb]
\includegraphics[clip,width=0.9\columnwidth,height=0.5\columnwidth]{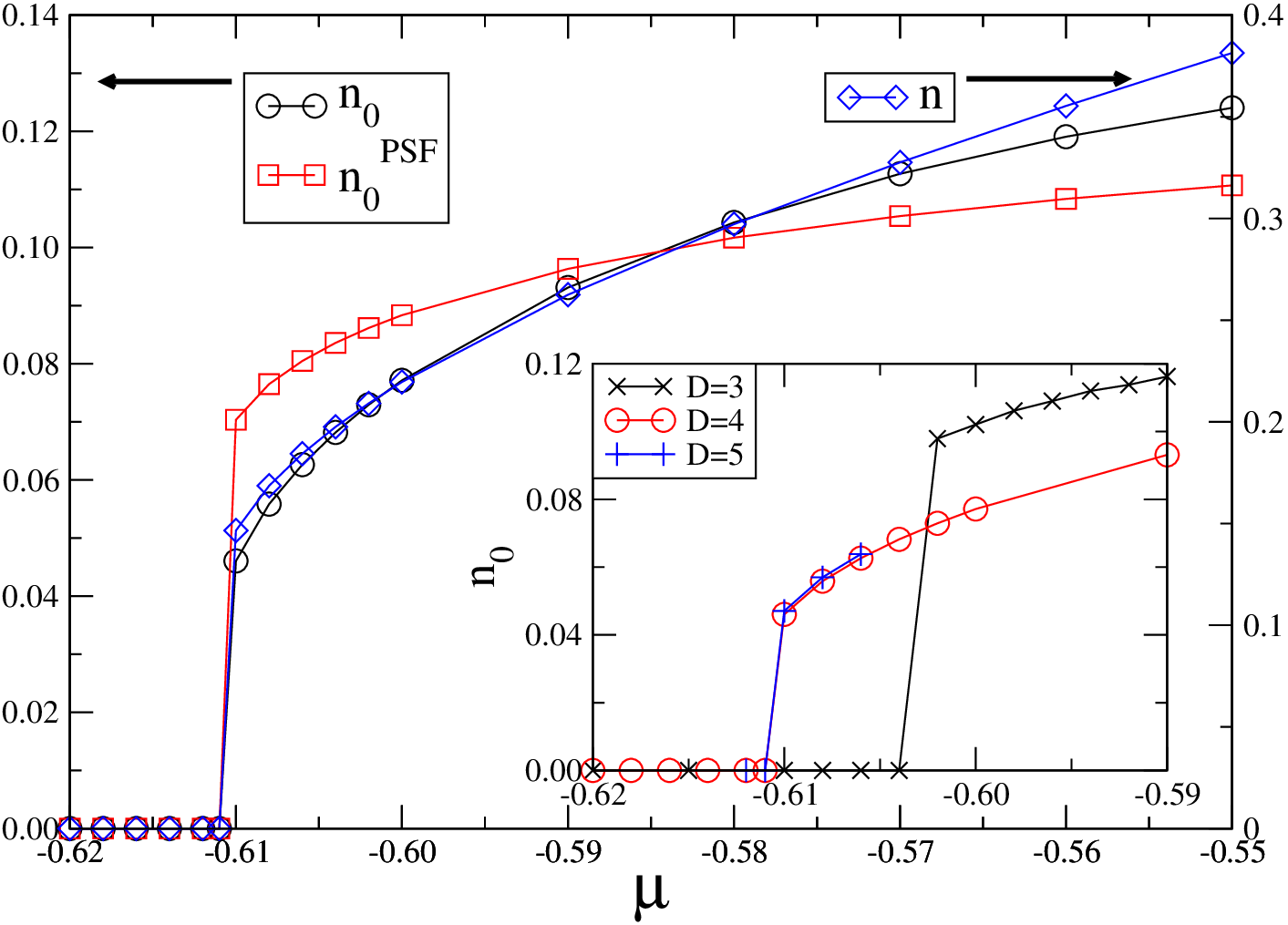}
\vskip 0.2cm
\includegraphics[clip,width=0.9\columnwidth,height=0.5\columnwidth]{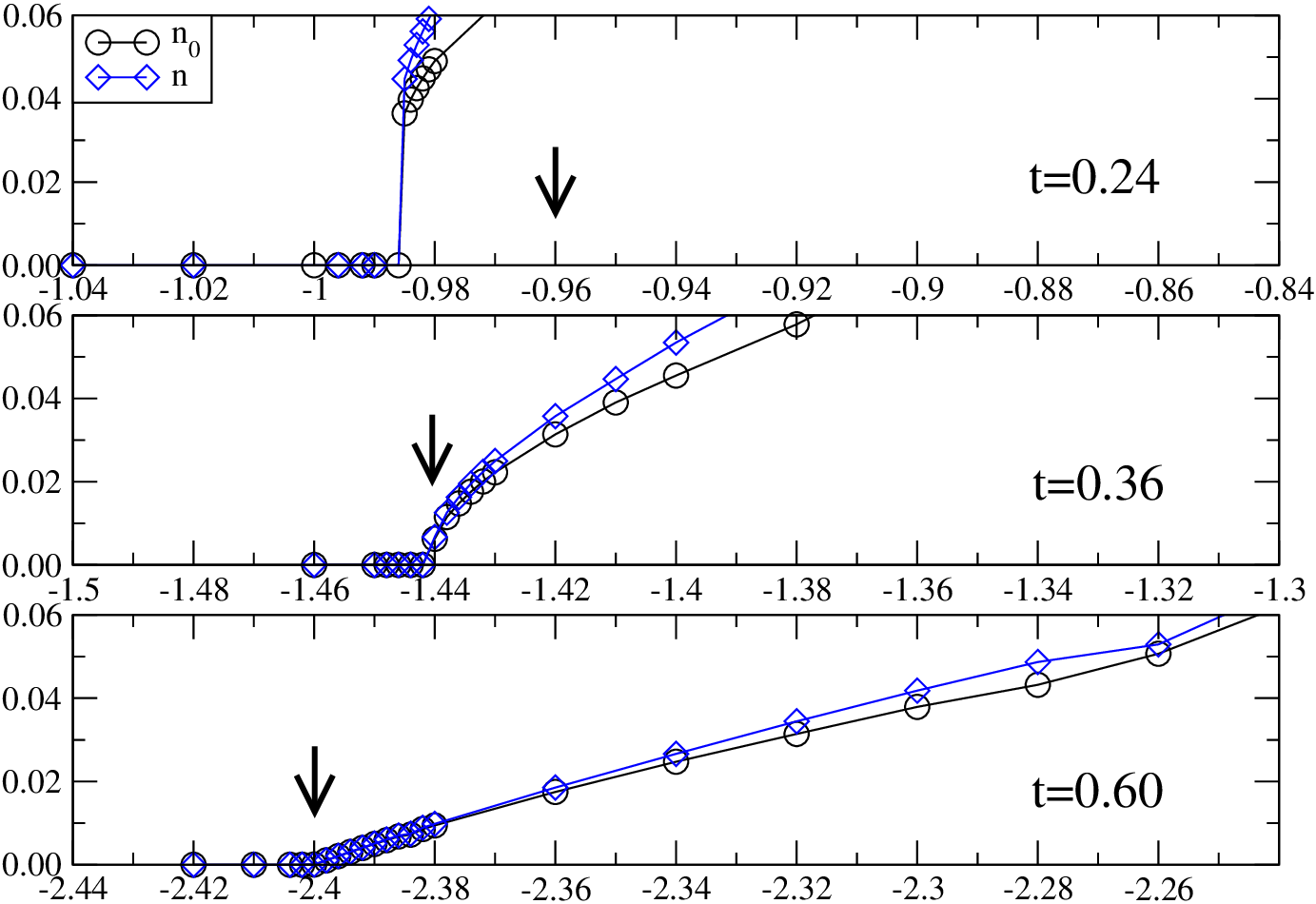}
\caption{(Color online) Upper panel: values of $n$, $n_0$ and
$n_0^\textrm{PSF}$ for the ground states at $t=0.12$ as functions
of chemical potential $\mu$. The inset shows the $D$ dependence of
$n_0$, where we use $D_{\rm cut}=16$ for $D=3$ and $D_{\rm
cut}=25$ for $D=4$, 5. Lower panel: $n$ and $n_0$ as function of $\mu$ at different $t$'s.
The arrows denote the analytic predictions of the transition
points, assuming that they are continuous transitions.
} \label{fig:scan1-2}
\end{figure}

These quantities are calculated under the combined iTEBD and TRG
algorithm~\cite{iTEBD-TRG}, where the ground-state wave function
is assumed in the form of the tensor product state (TPS) or the
projected entangled-pair state (PEPS)~\cite{TPS_review}. Due to
the similarity between the present systems and the spin-1/2
bilayer spin models, the same construction of TPS as that
discussed previously in Ref.~\cite{Chen10} for the study of the
bilayer systems is exploited here. Details can be found in
Refs.~\cite{Chen09,Chen10}. We note that the approximation in TPS
can be systematically improved simply by increasing the bond
dimension $D$ of the underlying tensors. The accuracy in
evaluating the expectation values for a TPS/PEPS ground state of
very large systems can be systematically improved by increasing
the TRG cutoff $D_{\rm cut}$. In this study, we consider the bond
dimension up to $D=5$ and keep $D_{\rm cut}\ge D^2$ to ensure the
accuracy of the TRG calculation. As shown in the insets of the
upper panels of Figs.~\ref{fig:scan3-4} and \ref{fig:scan1-2}, the
results for $D=4$ seem to converge to those for larger $D=5$. In
the following, we take $D=4$ and $D_{\rm cut}=25$ for systems with
size $N_s=2^7\times 2^7$ unless mentioned otherwise.

%%%%%%%%%%%%%%%%%%%%%%%%%%%%%%%%%%%%%%%%%%%%%%%%
%%%%%%%%%%%%%%%%%%%%%%%%%%%%%%%%%%%%%%%%%%%%%%%%%

%%%%%%%%%%%%%%%%%%%%%%%%%%%%%%%%%%%%%%%%%%%%%%%%%%%%%%%%%%%%%%%%%%
%\section{results}

The results of the local order parameters defined above as
functions of hopping parameter $t$ for chemical potentials
$\mu=-0.5$ and $-0.57$ are shown in Fig.~\ref{fig:scan3-4}. It is
found that, at half filling ($n=1/2$) with $\mu=-0.5$, there
exists a second-order quantum phase transition at $t = t_{c,{\rm
PSF-2SF}} \simeq 0.081$ between the PSF and the 2SF phases.
We note that, at half-filling, the particle-hole transformation
for a single species (say, $b_i \rightarrow b_i^\dag$) changes the
sign of interspecies interaction in Eq.~\eqref{eqn:H}, $U_{ab}
\rightarrow -U_{ab}$. Therefore, at this particular filling, the
results obtained for the attractive $U_{ab}$ model can be
reinterpreted in the context of the repulsive $U_{ab}$ model and
vice versa. In particular, the PSF order parameter $\langle ab
\rangle$ is mapped onto the super-counterflow (SCF) order
parameter $\langle ab^\dag \rangle$. Therefore, the critical value
of $t$ for the PSF-2SF transition of the attractive hard-core
boson model should be the same as the SCF-2SF transition of the
repulsive hard-core boson model studied
previously~\cite{altman03,soyler09}. We find that our critical
value of $t$ is consistent with the result ($t_{c,{\rm PSF-2SF}}
\simeq 0.092$) obtained by QMC method~\cite{soyler09}. This
indicates that the bond dimension used here is large enough to
provide accurate findings for the present model. As shown in
Fig.~\ref{fig:scan3-4}(a), within the PSF phase, the values of the
pair condensate density $n_{0}^{\rm PSF}$ do not remain the same
and are somewhat smaller than the constant value $n_{0}^{\rm PSF}=
1/4$ predicted by the mean-field theory~\cite{altman03}. This
reduction should result from quantum fluctuations therein.

While the PSF-2SF transition at $\mu=-0.5$ is of second order,
based on the mean-field analysis, it was suggested that this
transition could become first-order around the end point of the
PSF-2SF phase boundary~\cite{kuklov04-2}. If this conclusion is
true, the behavior of the first-order transition should become
more apparent as the end point is approached. In
Fig.~\ref{fig:scan3-4}(b), $n$, $n_0$, and $n_0^{\rm PSF}$ as
function of $t$ for $\mu=-0.57$ are displayed. A transition at
$t_{c,{\rm PSF-2SF}}\simeq 0.1024$ is found, beyond which $n_0$
becomes nonzero. Besides, anomalies in $n$ and $n_0^{\rm PSF}$
occur at this transition point. Although this transition point is
quite close to the end point of the PSF-2SF phase boundary as
indicated in Fig.~\ref{fig:phase_diag}, we find that the
transition is of second order. The same conclusions about the
order of transition are reached for larger $\mu$'s. Nevertheless,
we do observe that the order-parameter curves become more and more
first-order-like as $\mu$ decreases from $\mu=-0.5$ to $\mu=-0.57$
(not shown here). Our observation is consistent with the results
achieved by the Monte Carlo simulation on the related classical
model in Ref.~\cite{kuklov04-2}, where only the second-order
PSF-2SF transitions are found. However, one cannot completely rule
out the possibility of \emph{weakly first-order} transitions with
tiny jumps in order parameters when the transition points get much
closer to the end point of the PSF-2SF phase boundary.

As seen in Fig.~\ref{fig:scan3-4}(b), when $\mu=-0.57$, an
additional second-order phase transition between the $n=0$ MI and
the PSF states appears at $t_{c,{\rm MI-PSF}}\simeq 0.094$. In
this case, the insulating state is expected to be unstable when
the energy gap of pair excitation closes as $t$ increases. Above
the transition point, the lowest pair excitation starts to
condense and a PSF state develops. By using the expression of the
\emph{two}-particle excitation energies relative to the $n=0$
state up to the third-order perturbation expansion in
$t$~\cite{Iskin2010,note1}, the critical hopping parameter for
this transition is found to be $t_{c,{\rm MI-PSF}} =
\sqrt{(-2\mu-1)/4z}$, where $z=4$ is the coordination number for
square lattices. As shown in Fig.~\ref{fig:scan3-4}(b) and
Fig.~\ref{fig:phase_diag} for cases of other $\mu$'s, our findings
of $t_{c,{\rm MI-PSF}}$ agree well with the values given by the
analytical expression. This further substantiate that the employed
method can provide reliable results for the present system.

%%%%%%%%%%%%%%%%%%%%%%%%%%%%%%%%%%%%%%%%%%%%%%%%
%%%%%%%%%%%%%%%%%%%%%%%%%%%%%%%%%%%%%%%%%%%%%%%%%

When the single-particle hopping parameter $t$ is sufficiently
large, instead of the MI-PSF transition, a direct transition from
the $n=0$ MI state to the 2SF state as chemical potential $\mu$
being increased becomes possible. If this MI-2SF transition is of
second order, it should be induced by closing the energy gap of
\emph{single}-particle excitation. By using the analytical
expression of the one-particle excitation energies relative to the
$n=0$ state~\cite{Iskin2010,note1}, the critical value of $\mu$ at
fixed $t$ is given by $\mu_{c,{\rm MI-2SF}} = -zt$. However, as
discussed in Ref.~\cite{kuklov04-2}, the continuous MI-2SF
transitions can be preempted by the first-order ones when mutual
interaction of the dilute gapped quasiparticle excitations in the
MI phase is \emph{attractive}. Due to the energy gain from the
mutual attraction among quasiparticles, this first-order
transition can happen before closing one-particle excitation gap.
That is, the transition points for the first-order transitions
would be lower than those derived by the gap closing condition.
These discussions are confirmed by our calculations. In
Fig.~\ref{fig:scan1-2}, our results of the local order parameters $n$,
$n_0$ and $n_0^\textrm{PSF}$ as functions of $\mu$ for various $t$
are plotted. For $t=0.12$ (upper panel), a first-order MI-2SF
transition at $\mu_{c,{\rm MI-2SF}} \simeq -0.611$ is clearly
observed. Upon increasing $t$, the jumps in the order parameters
become smaller and smaller, and the transition eventually turns to
be of second order (see the lower panel). For example, the MI-2SF
transition at $t=0.6$ is found to be continuous, and it occurs at
$\mu_{c,{\rm MI-2SF}} \simeq -2.4$ in agreement with the analytic
formula. As shown in Fig.~\ref{fig:scan1-2}, the transition points
for those first-order transitions are always below the values
given by the gap closing condition. Thus the phenomena of the
transitions preempted by first-order ones is indeed demonstrated
in the present systems.

%%%%%%%%%%%%%%%%%%%%%%%%%%%%%%%%%%%%%%%%%%%%%%%%%%%%%%%%%%%%%%%%%%
%\section{conclusion}

To conclude, various quantum phase transitions of the two-species
hard-core boson model with attractive interspecies interaction are
explored under the combined iTEBD and TRG
algorithm~\cite{iTEBD-TRG}. Clear evidence of a PSF phase over a
finite regime in the phase diagram is provided. Critical values of
system parameters at corresponding first- or second-order phase
transitions are evaluated. Besides providing quantitative
predictions of the phase boundaries of the proposed PSF phase, the
present work also illustrates clearly the general validity and
flexibility in applying the current formalism to determine quantum
phase transitions in two dimensions.

%%%%%%%%%%%%%%%%%%%%%%%%%%%%%%%%%%%%%%%%%%%%%%%%%%%%%%%%%%%%%%
%\begin{acknowledgments}
We are grateful to K.-K. Ng and B. Capogrosso-Sansone for many
enlightening discussions.
P. Chen and M.-F.Yang thank the support from the NSC of Taiwan
under Contract No. NSC NSC 98-2112-M-007-010-MY3 and NSC
96-2112-M-029-004-MY3, respectively. This work is supported also
by NCTS of Taiwan.
%\end{acknowledgments}
%%%%%%%%%%%%%%%%%%%%%%%%%%%%%%%%%%%%%%%%%%%%%%%%%%%%%%%%%%%%%%

\end{document}